\documentstyle[prl,aps]{revtex}
\begin{document}
\twocolumn[
\title{Periodic Orbit Theory of Diffraction}
\author{G\'abor Vattay\cite{LAbs}}
\address{Niels Bohr Institute,
Blegdamsvej 17, DK-2100  Copenhagen \O, Denmark}
\author{Andreas Wirzba}
\address{Institut f\"ur Kernphysik, Technische Hochschule Darmstadt,
Schlo\ss gartenstr.\ 9,\\ D-64289 Darmstadt, Germany}
\author{Per E. Rosenqvist}
\address{Niels Bohr Institute,
Blegdamsvej 17, DK-2100  Copenhagen \O, Denmark}
\date{\today}
\maketitle

\mediumtext
\begin{abstract}
An extension of the Gutzwiller trace formula is given that includes
diffraction effects  due to hard wall scatterers or
other singularities. The new trace formula involves periodic orbits
which have arcs on the surface of  singularity and which correspond
to creeping waves. A new family of resonances in the two disk scattering
system can be well described  which is completely missing if only
the traditional periodic orbits are used.
\end{abstract}
] % end of \twocolumn
\narrowtext

Gutzwiller's trace formula\cite{Gut} is an increasingly popular tool for
analyzing semiclassical behaviour. Recently, it has been demonstrated
that using proper mathematical apparatus, like
Gutzwiller-Voros\cite{Voros} zeta functions, cycle expansions\cite{CE}
or Quantum Fredholm Determinants\cite{CV}, the trace formula can
successfully predict individual resonances of open scattering
systems\cite{Gaspard},
%g
it is possible to compute systematic $\hbar$ corrections\cite{hbar} and
to extend the formula for systems with regular domains\cite{Sieber}.
%g
The physical content of the trace formula is
the geometrical optical approximation of quantum mechanics via
canonical invariants of closed classical orbits. This approximation
is very accurate
when periodic orbits sufficently cover the phase space of the
chaotic system.
This is not the case when the number of
obstacles is small or their distance is large compared to their
typical size.  In such cases it is very important to take into account
the next-to-geometrical effects.

In this paper we study how the {\em
Geometric Theory of Diffraction} (GTD)
for hard core potentials
can be incorporated in the
periodic orbit theory.
Such a problem
occurs where the  wave length of a quantum mechanical
(or optical) wave is very large compared with the spatial variation of
a repulsive potential, e.g.\ at the boundaries of
microwave guides, optical fibers, superconducting squids, or circuits in the
ballistic evolution of electrons,
i.e.\ in most of the devices used for so-called macroscopic
quantum mechanical (or optical) experiments.
First we summarize from the classical
papers of J.~B.~Keller\cite{Keller1} how the
Green's function in the shadowed regions of configuration space can be
computed by the concept of GTD. Then we incorporate the periodic rays
with diffracted ray arcs to the trace formula.
%g
It is possible to derive the diffraction part of the Green's function
directly from the semiclassical approximation of the Feynman path
integral in the neighborhood of a hard curved wall\cite{Buslaev}.
We have choosen Keller's approach here, since it is more suitable for
computing the trace of the Green's function, without further approximations.

It has been known  for quite a long time that the scattering amplitude
in the shadowed region behind an obstacle can be well reproduced by
allowing diffracted rays in addition to geometrical ones\cite{Keller1}.
The diffracted rays
connecting two points in the configuration space in the presence of sharp
objects can be derived from an extension of  Fermat's variational principle
of classical mechanics\cite{Keller1}.
The usual Fermat principle states that the classical trajectories
connecting two positions $q_{\cal A}$ and $q_{\cal B}$ in
configuration space are those smooth curves which make the action
stationary.
If the configuration
space is bounded by hard walls, a generalized
variational principle, introduced by J.~B.~Keller\cite{Keller1},
has to be applied.
This principle requires new classes of curves:
We have to consider for each triplet
of integers $r,s,t\ge 0$ the class of curves ${\cal D}_{rst}$
with $r$ smooth arcs on the surface, $s$ points on the edges
and $t$
points on the vertices of the boundary or the discontinuity. The curves
of the GTD are those which make the action stationary within
one of the classes ${\cal D}_{rst}$. The class ${\cal D}_{000}$ corresponds
to the usual geometrical orbits.
In this paper, we concentrate on
two dimensional problems, where the simplest nontrivial curves
are of class  ${\cal D}_{100}$ and ${\cal D}_{001}$
(Fig.~1a and b respectively),
whereas edge diffraction ${\cal D}_{010}$ is not possible.
Once we know the generalized ray we can compute semiclassically the
Green's function $G(q_{\cal A},q_{\cal B},E)$ tracing the ray
along it\cite{Keller1}. In Fig.\ 1a (${\cal D}_{100}$)
the  trajectory -- obtained from the generalized variational principle -- is
tangent  to the surface of the hard wall obstacle at the points $\cal A'$
and $\cal B'$. The Green's function
$G(q_{\cal A},q_{\cal A'},E)$ can be
computed semiclassically by the energy domain Van-Vleck propagator (for a
single classical trajectory)
\begin{equation}
G(q,q',E)=
\frac{2\pi}{(2\pi i \hbar)^{3/2}}D_{\text{V}}^{1/2}(q,q',E)
e^{\frac{i}{\hbar}S(q,q',E)-\case{i}{2}\nu\pi},
\end{equation}
where $D_{\text{V}}(q,q',E)=|\det(-\partial^2 S/\partial q_i
\partial q_j')|/|\dot{q}||\dot{q}'|$ is the Van-Vleck determinant
and $\nu$ is the Maslov index (see ref.\cite{Gut3} for definitions).

When the geometrical ray hits the surface of the obstacle it creates a
source for the diffracted (creeping)
wave. The strength of the source is proportional to
the Green's function at the incidence of the ray
\begin{equation}
Q_{\text{diff}}=DG_{\text{inci}}\ .\label{D}
\end{equation}
The diffraction constant $D$ depends on the local geometry and
the nature of the diffraction. It has been determined
in Ref\cite{Keller1} from the asymptotic
semiclassical expansion of the exact solution in some
simple geometry\cite{Keller1} (see also \cite{Franz}). Its form is
\begin{equation}
D_l= 2^{{1}/{3}} 3^{{-2}/{3}}\pi e^{5 i {\pi}/{12}}
\frac{(k \rho)^{{1}/{6}}}{Ai'(x_l) }\ .\label{DL}
\end{equation}
Here  $Ai'(x)$ is the derivative of the Airy integral
$Ai(x)=\int_{0}^{\infty}dt\cos(xt-t^3) $, $k=\sqrt{2mE/\hbar}$
is the wave number, $\rho$ is the
radius of the obstacle at the source of the creeping ray and
$x_l$ are the zeroes of the Airy integral, which can
be approximated as
$
x_l= 6^{{1}/{3}}\,(3\pi(l-1/4))^{2/3}/2
$
in the semiclassical limit.
The index $l\geq 1$ refers to
the possibility of initiating creeping rays with different modes,
each with its own profile. In practice only the low modes
contribute to the Green's function.
The source then initiates a ray creeping along the surface.
During the creeping  of the ray the amplitude decreases,
which can be understood as a process analogous to the
radiation processes of electrodynamics.
The radiated intensity is proportional to the intensity of the
ray:
\begin{equation}
\frac{d}{ds}A_l(s,E)^2 =-2  \alpha_l(s,E)\, A_l(s,E)^2,
\end{equation}
where $s$ is the length measured along the surface and $A_l(s,E)$
is the complex amplitude of the Green's function along the surface.
The coefficient $\alpha_l(s,E)$ depends on the local curvature of
the surface, $1/\rho(s)$, and it has the structure
$
\alpha_l(s,E)=x_l e^{-i{\pi}/{6}}({k}/{6\rho(s)^2 })^{{1}/{3}}
$ (see Ref.\cite{SS}),
where the index $l$ refers again to the different modes of the
creeping wave.
The Green's function for the creeping ray of mode $l$ is then given by
\begin{equation}
G^{D}_l(q_{\cal A'},q_{\cal
B'},E)=e^{-\int_0^{L}ds\alpha_l(s,E)}e^{\frac{i}{\hbar}S(q_{\cal
A'},q_{\cal B'},E)},\label{Gcr}
\end{equation}
where $L$ is the length of the arc of the creeping ray,
and $S(q_{\cal A'},q_{\cal B'},E)$ is the  action along it.
The creeping ray at the point ${\cal B'}$ initiates a pure geometrical
ray. The source of this ray is located in ${\cal B'}$, and its strength is
again given by equations (\ref{D}) and (\ref{DL}) due to the
invariance of the Green's function against the interchange of the
variables $q_{\cal A'}$ and $q_{\cal B'}$.
The total Green's function is then the product of the Green's functions
and diffraction coefficients along the ray:
\begin{eqnarray}
G(q_{\cal A},q_{\cal B},E)&=& G(q_{\cal A},q_{\cal
A'},E)\sum_{l=1}^{\infty}D_{l,{\cal A'}}G^{D}_l(q_{\cal A'},q_{\cal
B'},E)  \nonumber \\
& & \times
D_{l,{\cal B'}}  G(q_{\cal B'},q_{\cal B},E).
 \label{Gprod}
\end{eqnarray}
In a general situation, when the ray consists of several pure geometric
and creeping arcs, the Green's function can also be written
as a product of partial Green's functions and diffraction constants.

To incorporate diffraction effects into the trace formula, one should
compute the trace of the Green's function derived above. As in the case
of the Gutzwiller trace formula -- derived from a pure geometrical
approximation of the Green's function -- the trace receives the leading
contributions from tubes encircling the closed curves, which
now can have diffractional arcs too.
We can handle separately the pure geometric cycles and the cycles with
{\em at least one diffraction arc} along one of the obstacles:
\begin{equation}
\text{Tr}\:  G(E) \approx \text{Tr}\:  G_{G}(E) + \text{Tr}\:  G_{D}(E),
\end{equation}
where $\text{Tr}\:  G_{G}(E)$ is the ordinary Gutzwiller trace formula,
while $\text{Tr} \:   G_{D}(E)$
is the new trace formula corresponding
to the non-trivial cycles
of the GTD.
$\text{Tr}\:  G_{D}(E)$ can be computed by using appropriate
Watson\cite{Watson,Franz} contour integrals. For technical details we
invite the reader to Refs.\cite{Andreas1} and\cite{Andreas2}. Here we
communicate the general result and the detailed calculation will be
published elsewhere\cite{VWR2}. If we denote by $q_i$, $i=1,\dots,n$
(with $q_{n+i}\equiv q_i$) the
points along
the closed cycle, where the ray  changes from diffraction to pure geometric
evolution or vice versa (see ${\cal A'}$ and ${\cal B'}$ in Fig.\
1a),
the trace for cycles with
{\em at least one diffractional arc}
can be expressed as the product
\begin{equation}
\text{Tr}\:  G_{D}(E)=\sum_{\text{Cycles}}\frac{T(E)}{i\hbar}\prod_{i=1}^{n}
                              {D}(q_i)G(q_i,q_{i+1},E),
\end{equation}
where $T(E)$ is the time period of the cycle (without repeats)
and $D(q_i)$
is the diffraction constant (\ref{DL}) with the radius of curvature $\rho$
given
locally at the point $q_i$. The creeping mode index $l$ and the
corresponding summations
(see e.g.\  (\ref{Gprod}))
are surpressed here for keeping the notation simple.
$G(q_i,q_{i+1},E)$ is
alternatingly
the Van-Vleck propagator, if $q_i$ and $q_{i+1}$ are connected by pure
geometric
arcs, or is given by (\ref{Gcr}) in case $q_i$ and $q_{i+1}$ are the boundary
points of  a creeping arc.
Note this formula does only apply for cycles with
at least one creeping section. Such cycles have the special property
that their pertinent energy domain Green's functions
are multiplicative (see e.g.\ (\ref{Gprod}) with the summations over the
creeping
mode numbers of course included consistently). This does not
hold for pure geometrical cycles.

The eigenenergies can be recovered from the Gutzwiller-Voros spectral
determinant\cite{Voros} $\Delta(E)$, which is related to the trace formula as
\begin{equation}
\text{Tr}\:  G(E)=\frac{d}{dE}\ln \Delta(E).
\end{equation}
The full semiclassical determinant can be written as the
{\em formal}
product of
two spectral determinants, one corresponding to pure geometrical
and one to new cycles
$\Delta(E)=\Delta_G(E)\Delta_D(E)$ due to the additivity of the
traces.
The product is only formal, since the eigenenergies
are not given by the zeros of $\Delta_G(E)$ or $\Delta_D(E)$ individually, but
have to be calculated from a curvature expansion of the {\em combined}
determinant
$\Delta (E)$ itself.

The diffraction part of the spectral determinant is
\begin{equation}
\Delta_D(E)=\exp\left(-\sum_{p,r=1}^{\infty}\frac{1}{r}\prod_{i=1}^{n_p}
[ D(q^p_i)G(q^p_i,q^p_{i+1},E)]^r\right),
\end{equation}
where the summation goes over closed primitive (non-repeating)
cycles $p$ and the repetition number $r$.
The product of Green's functions should be evaluated for
$q^p_i$ belonging to the primitive cycle $p$.
After summation over $r$, the spectral determinant can be written as
\begin{equation}
\Delta_D(E)=\prod_p(1-t_p) \label{Dp}
\end{equation}
with
\begin{equation}
t_p=\prod_{i=1}^{n_p} { D}(q^p_i)G(q^p_i,q^p_{i+1},E),
\label{We}
\end{equation}
where $q^p_i$ belongs to the primitive cycle $p$.
As mentioned before
the mode numbers $l$ of the diffraction constants and the corresponding
summations have been surpressed for notational simplicity;
they can be easily restored
as e.g.\ in the final expression (\ref{final}).

We can conclude that the  diffractional part $\Delta_D(E)$ of the spectral
determinant shares some nice features of the periodic orbit
expansion of the dynamical zeta functions\cite{CE},
and it can be expanded as
\begin{equation}
\Delta_D(E)=1-\sum_p t_p +\sum_{p,p'}t_p t_{p'}- \cdots .
\end{equation}
The weight (\ref{We}) has the following property  which helps in
radically reducing the number of relevant contributions in the
expansion:
If two different cycles $p$ and $p'$ have at least one common piece in their
diffraction arcs, then the two cycles can be composed to one
longer cycle $p+p'$ and the weight corresponding to this longer cycle
is the product of the weights of the short cycles
\begin{equation}
t_{p+p'}=t_{p}\cdot t_{p'}.
\end{equation}
As a consequence, the product of primitive cycles, which have
at least one common piece in their diffraction arcs, can be
reduced in such a way that the composite cycles are exactly
cancelled in the curvature expansion
\begin{equation}
\prod_p(1-t_p)=1-\sum_b t_b,
\end{equation}
where $t_b$ are {\em basic} primitive orbits
which can not be composed from shorter primitive
orbits.

To demonstrate the importance of the diffraction effects
to the spectra, we have calculated the $A_1$  resonances of
the scattering system of two equally sized hard circular disks with
disk separation $R=6a$, where $a$ is the radius of one disk.
In this system there is only
one geometrical periodic cycle along the line connecting
the centers of the disks. Its stability
$\Lambda_p=9.8989794$ and action $S_p=kL_p=k\cdot4 a$ yield the
geometrical part of the spectral determinant\cite{CE2,Andreas2}
\begin{equation}
\Delta_G(k)=\prod_{j=0}^{\infty}
 \left(1+\frac{e^{ikL_p}}{\Lambda_p^{(1+4j)/2}}\right), \label{dgeo}
\end{equation}
where $k=\sqrt{2mE}/\hbar$ and $2m=\hbar=1$, and  leads to the following
predictions for the semiclassical $A_1$ resonances
\begin{equation}
k^{\text{res}}_{n,j}=( \pi(2n-1)-i\case{1+4j}{2}\,\ln \Lambda_p )/L_p
\label{resgeo}
\end{equation}
with $n=1,2,3,\cdots$.
Note in the above expressions $(1+4j)/2$ replaces the
usual weight $(1+2j)/2$,
since the geometrical orbit in the two-disk problem lies on the boundary of the
fundamental domain \cite{CE2}.

Fig.\ 1c shows the
first four new basic cycles in the fundamental
domain\cite{CE2}.
We computed
the geometrical data of the first ten orbits and used them to construct
the creeping and geometrical Green's functions.
The semiclassical Green's function in free space is asymptotically
($ k R \gg 1$)
\begin{equation}
G_0(q,q',E)=-\frac{i}{4}\left(\frac{2}{\pi
kR}\right)^{1/2}e^{ikR-i\case{\pi}{4}},
\label{GRB}
\end{equation}
where $R=|q-q'|$.
If the ray connecting $q$ and $q'$ is reflected once or more
from the curved hard walls  before hitting tangentially one of the
surfaces, we can keep track of the change in the amplitude by the help of the
Sinai-Bunimovich curvatures.

By computing the curvatures $\kappa_i$ right after the reflections,
and knowing the distances $l_i$ between the $i$-th and the
$(i+1)$-th points of reflections, the factor $R$ in the Green's function
(\ref{GRB})
has to be changed to the effective radius
$R^{\text{eff}}=R_0\prod_{i=1}^{m}(1+l_i\kappa_i)$ where $R_0$ is the
distance between $q$ and the first point of reflection along the ray starting
from $q$, and $m$ is the number of reflections from a disk.
The effective radius $R_b^{\text{eff}}$, the length of the geometrical
arc  $L_b^G$ and
the length of the diffraction part $L_b^D$ of the first ten orbits with
creeping sections are listed in Table \ref{tab}.
To each cycle in the list, there is a whole sequence
of cycles which wind around the disk $m_{\text{w}}$ times. For these
orbits one has to add $2\pi a m_{\text{w}}$ to the diffraction length $L^D_b$.
The diffraction part of the spectral determinant is
finally given by
\begin{eqnarray}
\Delta_D(k)&=&1-\sum_{b,l}
(-1)^{m_b}C_l\frac{a^{{1}/{3}}e^{i{\pi}/{12}}e^{ik(L^G_b+L^D_b)-\alpha_l
L^D_b}}{k^{{1}/{6}}\sqrt{R_b^{\text{eff}}}} \nonumber \\
& & \times
\frac{1}{1-e^{2\pi (ik-\alpha_l)a}},
 \label{final}
\end{eqnarray}
where $C_l=\pi^{3/2}3^{-4/3}2^{-5/6}/Ai'(x_l)^2$, and the summation
for the windings $m_{\text{w}}$ gives the factor
$1/(1-e^{2\pi(ik-\alpha_l)a})$.  We computed the spectra by truncating
the product $\Delta_G(k)\Delta_D(k)$ at maximal cycle length 5 and
using only the $l=1$ term
in the now restored summation over the creeping mode number.
The exact quantum mechanical resonances were computed following
Ref.\cite{Andreas1}.

The leading semiclassical resonances are given
equally well with and without creeping
modifications.
In Fig.\ 2 we can see that the new formula
describes the resonances of the two disk system with a few-percent
error, while the computation based on the geometrical cycle alone,
Eq.\ (\ref{dgeo}), gives completely false results for the next-to-leading
resonances (see Eq.(\ref{resgeo})).

The authors are very grateful to  P. Cvitanovi\'c for encouraging
the project.
A.W.\ thanks NORDITA and NBI
for hospitality during his stay.
G.V.\ and P.E.R.\ thank  SNF for support
and G.V.\ acknowledges the support of the Sz\'echenyi Foundation, Phare
Accord H 9112-0378
and OTKA F4286.

\begin{table}
\caption{Table of the first ten basic cycles $t_b$ which include
creeping sections in the fundamental region of two-disk problem (with
disk separation $R=6a$). The
cycles are labelled by their number $m_b$ of geometrical reflections
from one of the disks. The length of the geometrical arc $L_b^G$, the
effective radius $R_b^{\text{eff}}$ and the length of the diffraction
section $L_b^D$ are listed in units of the disk radius $a$.\label{tab}}
\begin{tabular}{c c c c}
$m_b$ & $L_b^G/a$ & $R_{b}^{\text{eff}}/a$ & $L_b^D/a$ \\ \hline
0& 5.656854249492&   5.656854249424&   3.821266472498\\
0& 6.000000000000&   6.000000000000&   3.141592653589\\
1& 9.832159566199&   58.16784043380&   3.476488812029\\
1& 9.797958971132&   58.78775382679&   3.544308495170\\
2& 13.81654759452&   578.1406653460&   3.507404058891\\
2& 13.81309379078&   579.7434283719&   3.514253447057\\
3& 17.81499162871&   5729.649817456&   3.510488616089\\
3& 17.81464272590&   5732.235502463&   3.511180541615\\
4& 21.81483475355&   56728.70010470&   3.510799703655\\
4& 21.81479950722&   56732.26871144&   3.510869602322\\
\end{tabular}

\end{table}

\begin{figure}
\caption{The simplest classes ${\cal D}_{100}$ (a) and ${\cal D}_{001}$ (b)
of curves in two dimensions.
In the window (c): the first four basic orbits in the fundamental domain of
the two-disk system.}
\end{figure}

\begin{figure}
\caption{Resonances for the $A1$ subspace of the two-disk system
(with disk separation $R=6a$) in the complex $k$ plane in units of
the disk radius $a$.
The diamonds label the exact quantum mechanical resonances, which are
the poles of the scattering matrix. The crosses are their semiclassical
approximations including the diffraction terms derived in this paper.
The boxes
refer to the ordinary Gutzwiller semiclassical approximation
(eq.(17), $j=0,1$),
where the diffraction effects are not included.}
\end{figure}

\begin{references}
\bibitem[*]{LAbs} On leave from E\"otv\"os University, Budapest, Hungary.
\bibitem{Gut} M. C. Gutzwiller, J. Math. Phys. {\bf 12}, 343 (1971)
\bibitem{Voros} A. Voros, J. Phys. {\bf A21}, 685 (1988)
\bibitem{CE} P. Cvitanovi\'c, B. Eckhardt, Phys.Rev.Lett.{\bf 63}, 823 (1989)
\bibitem{CV} P. Cvitanovi\'c and G. Vattay, Phys. Rev. Lett. {\bf 71},
4138, (1993), P. Cvitanovi\'c, P. E. Rosenqvist, G. Vattay and H. H.
Rugh, CHAOS {\bf 3} in press (1993)
\bibitem{Gaspard} P. Gaspard and S.A. Rice, J. Chem. Phys. {\bf 90}, 2225
(1989); {\bf 90}, 2242 (1989); {\bf 90}, 2255 (1989)
\bibitem{hbar} P. Gaspard, D. Alonso, Phys. Rev. {\bf A47}, R3468, (1993)
\bibitem{Sieber} M. Sieber, U. Smilansky, S. C. Creagh and R. G. Littlejohn,
J. Phys. {\bf A26}, 6217, (1993)
\bibitem{Miller} G. E. Zahr and W. H. Miller, Mol. Phys. {\bf 30}, 951 (1975)
\bibitem{Buslaev} V. B. Buslaev, Sov. Phys. Doklady {\bf 7} 685 (1963),
E. H. Lieb, J. Math. Phys. {\bf 8}, 43 (1967)
L. S. Schulman, {\em Techniques and Applications of Path
Integration}, John Wiley \& Sons, New York (1981), pp. 333-339
\bibitem{Keller1} J. B. Keller, J. Opt. Soc. Amer. {\bf 52} 116 (1962)
%g \bibitem{Keller2}
J. B. Keller, in {\em Calculus of Variations
and its Application} (American Mathematical Society, 1958) p.27
%\bibitem{Gut2} W. Pauli {\em Feldquatisierung} (Techn. Hochschule in
%Z\"urich, 1950) p. 139, Ph. Choquard, Helv. Phys. Acta {\bf 28}, 89
%(1955), V. B. Busalev, Sov. Phys. Doklady {\bf 7} 685 (1963),
%M. C. Gutzwiller, J. Math. Phys. {\bf 8}, 1979 (1967), {\bf 10}, 1004 (1969),
%{\bf 11}, 1791 (1971)
%\bibitem{Bus} V.B. Busalev, Sov. Phys. Doklady {\bf 7} 685 (1963), {\bf 10}
%17 (1965)
%g \bibitem{Keller3}
B. R. Levy and J. B. Keller, Comm. Pur. Appl. Math.
{\bf XII}, 159 (1959)
%g \bibitem{Keller4}
B. R. Levy and J. B. Keller, Cann. J. Phys. {\bf 38}, 128
(1960)
\bibitem{Franz} W. Franz, {\em Theorie der Beugung Elektromagnetischer
Wellen}, Springer Verlag, Berlin (1957); Z. Naturforschung {\bf 9a}, 705
(1954)
\bibitem{Gut3} M. C. Gutzwiller, {\em Chaos in classical and quantum
mechanics},
Springer Verlag, New York (1990)
\bibitem{SS} B. Schrempp, F. Schrempp, Nucl.Phys.{\bf B163},397 (1980)
\bibitem{Watson} G. N. Watson, Proc. R. Soc. London {\bf A95}, 83 (1918)
\bibitem{Andreas1} A. Wirzba, CHAOS {\bf 2}, 77, (1992)
\bibitem{Andreas2} A. Wirzba, Nucl. Phys. {\bf A560}, 136, (1993)
\bibitem{VWR2} G. Vattay, A. Wirzba and P. E. Rosenqvist, in prep.
\bibitem{CE2} P. Cvitanovi\'c, B. Eckhardt, Nonlinearity {\bf 6},
277, (1993)
\end{references}
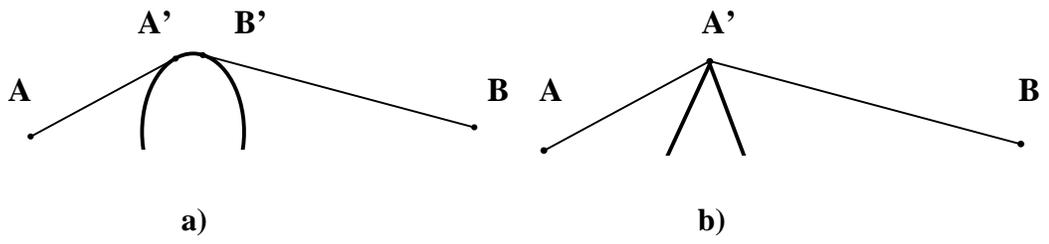
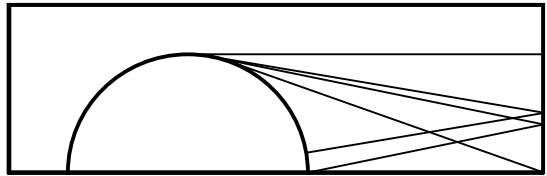
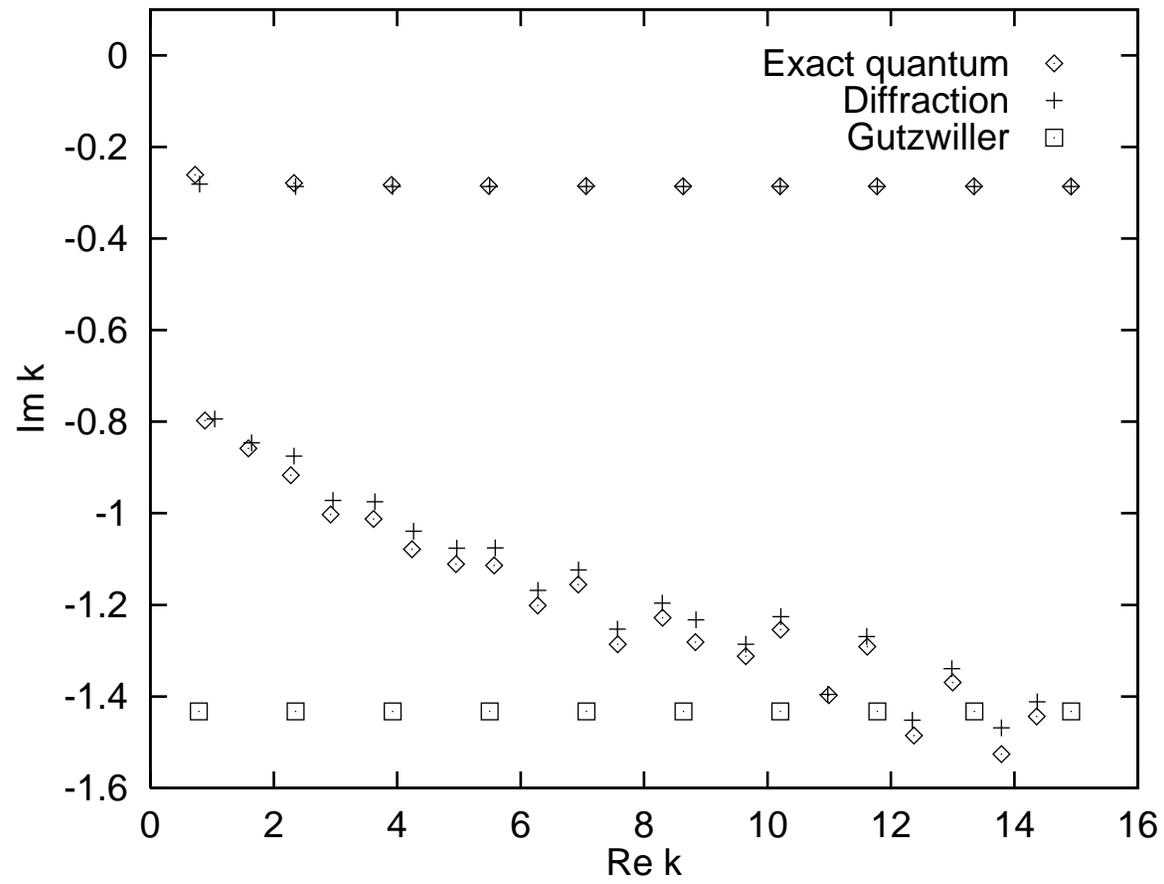
\end{document}